\documentclass[a4paper]{jpconf}
\usepackage{graphicx}
\usepackage{amsmath}
\usepackage{amssymb}
\newcommand{\be}{\begin{equation}}
\newcommand{\ee}{\end{equation}}

\newcommand {\vmin}{\mbox{$v_{min}$}}

\newcommand{\Ed}{E^{\prime}}
\newcommand{\ud}{\text{d}}

\newcommand{\eR}{\mathcal{R}}
\begin{document}
\title{Is a WIMP explanation of the DAMA modulation effect still viable?}

\author{Gaurav Tomar$^a$, Sunghyun Kang$^a$, Stefano Scopel$^a$, Jong-Hyun Yoon$^b$}

\address{$^a$Department of Physics, Sogang University, Seoul, Korea, 121-742}
\address{$^b$Department of Physics, University of Helsinki, FI-00014 Helsinki, Finland}
\ead{tomar@sogang.ac.kr}

\begin{abstract}
We show that the weakly interacting massive particle (WIMP) scenario of proton-philic spin-dependent inelastic dark matter can still provide a viable explanation of the observed DAMA effect in compliance with the constraints from other experiments. We also show that, although the COSINE-100 collaboration has recently tested the DAMA effect using the same target material, for the time being the comparison between DAMA and COSINE-100 still depends on the particle-physics model.
\end{abstract}

\section{Introduction}
The DAMA collaboration has been measuring for more than 15 years a yearly modulation effect in their sodium iodide target. Such effect has a statistical significance of more than 9$\sigma$ and is consistent with what is expected from dark matter (DM) WIMPs. Recently, the DAMA/LIBRA-phase2 result has been released where, compared to previous data, now the energy threshold has been lowered from 2 keV electron-equivalent (keVee) to 1 keVee and the exposure has almost doubled~\cite{Bernabei:2018yyw}. As a result of lower threshold, DAMA/LIBRA experiment is now sensitive to WIMP-iodine interactions at low WIMP masses. In this work, we discuss a phenomenological scenario, proton-philic spin-dependent inelastic dark matter (pSIDM), that we have shown to explain the DAMA effect in agreement with the constraints from other experiments. Since the COSINE-100 collaboration has employed a similar target ({\it NaI}) as DAMA and recently published their result for the spin-independent isoscalar WIMP-nucleus interaction~\cite{Adhikari:2018ljm}, we tested the pSIDM scenario against the COSINE-100 result.

\section{The pSIDM Scenario}
There are stringent bounds on  an interpretation of the DAMA effect in terms of WIMP-nuclei scattering coming from XENON1T, CDMS and PICO-60 experiments. While the spin of xenon and germanium is mostly originated by an unpaired neutron, the spin for both sodium and Iodine is due to an unpaired proton. This implies that if the WIMP particle interacts with ordinary matter predominantly via a spin-dependent coupling which is suppressed for neutrons, it can explain the DAMA effect in compliance with xenon and germanium bounds. To achieve this, in the pSIDM scenario, we tuned the neutron to proton coupling ratio $c^n/c^p=-0.028$. But this scenario is still constrained by droplet detectors and bubble chambers (COUPP, PICASSO, PICO-60) which all use nuclear targets with an unpaired proton ($^{19}F$ and/or $^{127}I$).

Through inelastic DM (IDM) it is possible to reconcile the above scenario. In IDM a DM particle $\chi_1$ of mass 
$m_{\chi_1}=m_{\chi}$ interacts with atomic nuclei exclusively by up--scattering to a second heavier
state $\chi_2$ with mass $m_{\chi_2}=m_{\chi}+\delta$. Basically, a minimal WIMP incoming speed is needed in the
lab frame matching the kinematic threshold for inelastic upscatters
given by,

\begin{equation}
v_{min}^{*}=\sqrt{\frac{2\delta}{\mu_{\chi N}}},
\label{eq:vstar}
\end{equation}

\noindent with $\mu_{\chi N}$ the WIMP--nucleus reduced
mass. If the WIMP mass $m_\chi$ and the mass-splitting $\delta$ are chosen in such a way  
that the hierarchy between $v^*_{min}$ for sodium $v_{min}^{*Na}$ and for fluorine $v_{min}^{*F}$ with the WIMP escape velocity 
$v_{esc}$,

\begin{equation}
v_{min}^{*Na}<v_{esc}^{lab}<v_{min}^{*F},
\label{eq:hierarchy}
\end{equation}

\noindent is achieved then WIMP scatterings off
fluorine turn kinematically forbidden while those off sodium can
still serve as an explanation to the DAMA effect. Clearly, the trivial observation that the velocity
$v_{min}^*$ for fluorine is larger than that for sodium is at the core of the pSIDM
mechanism.
\section{Analysis}
\begin{figure}
\begin{center}
\includegraphics[width=0.451\columnwidth]{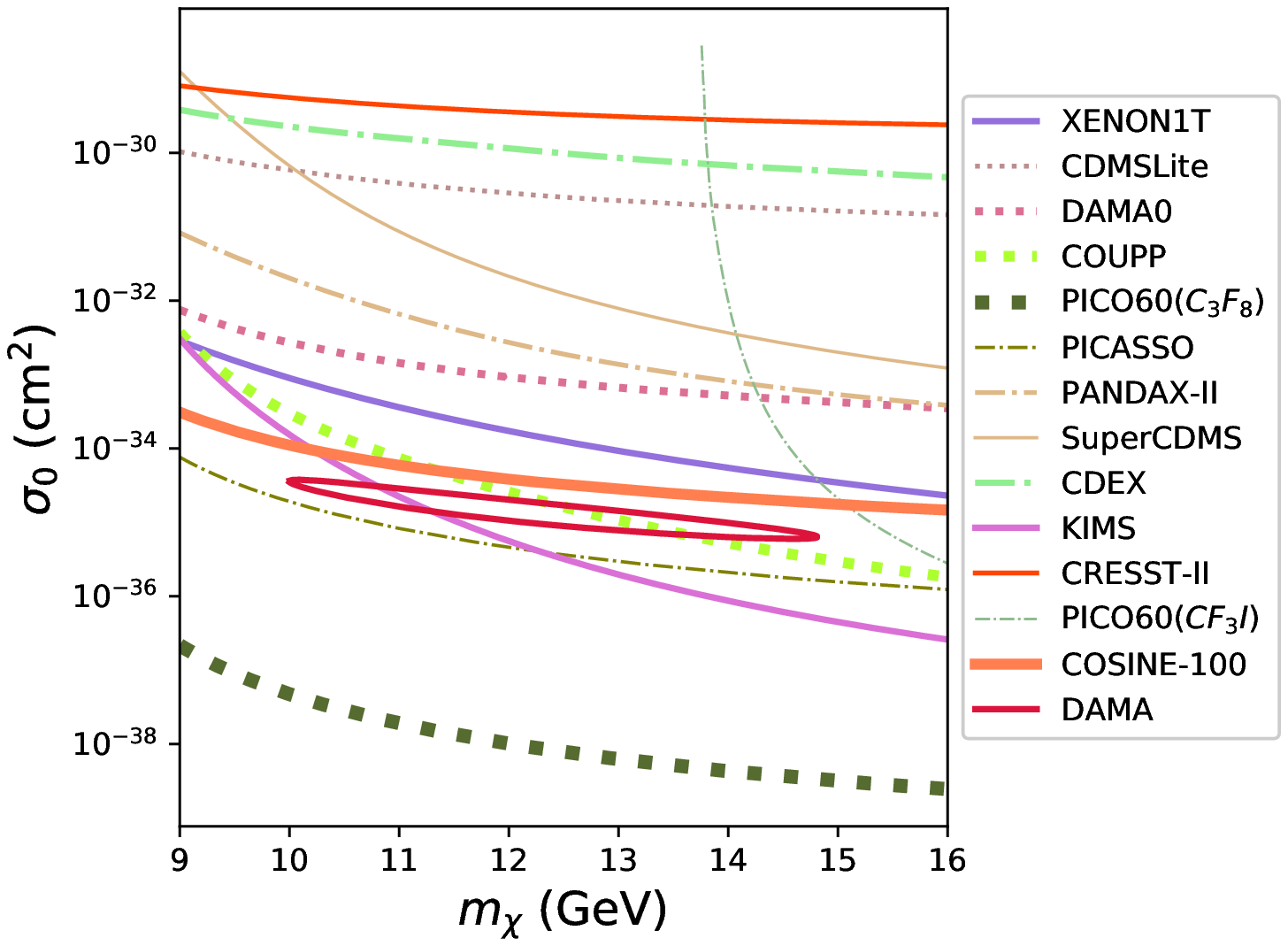}    
\includegraphics[width=0.49\columnwidth]{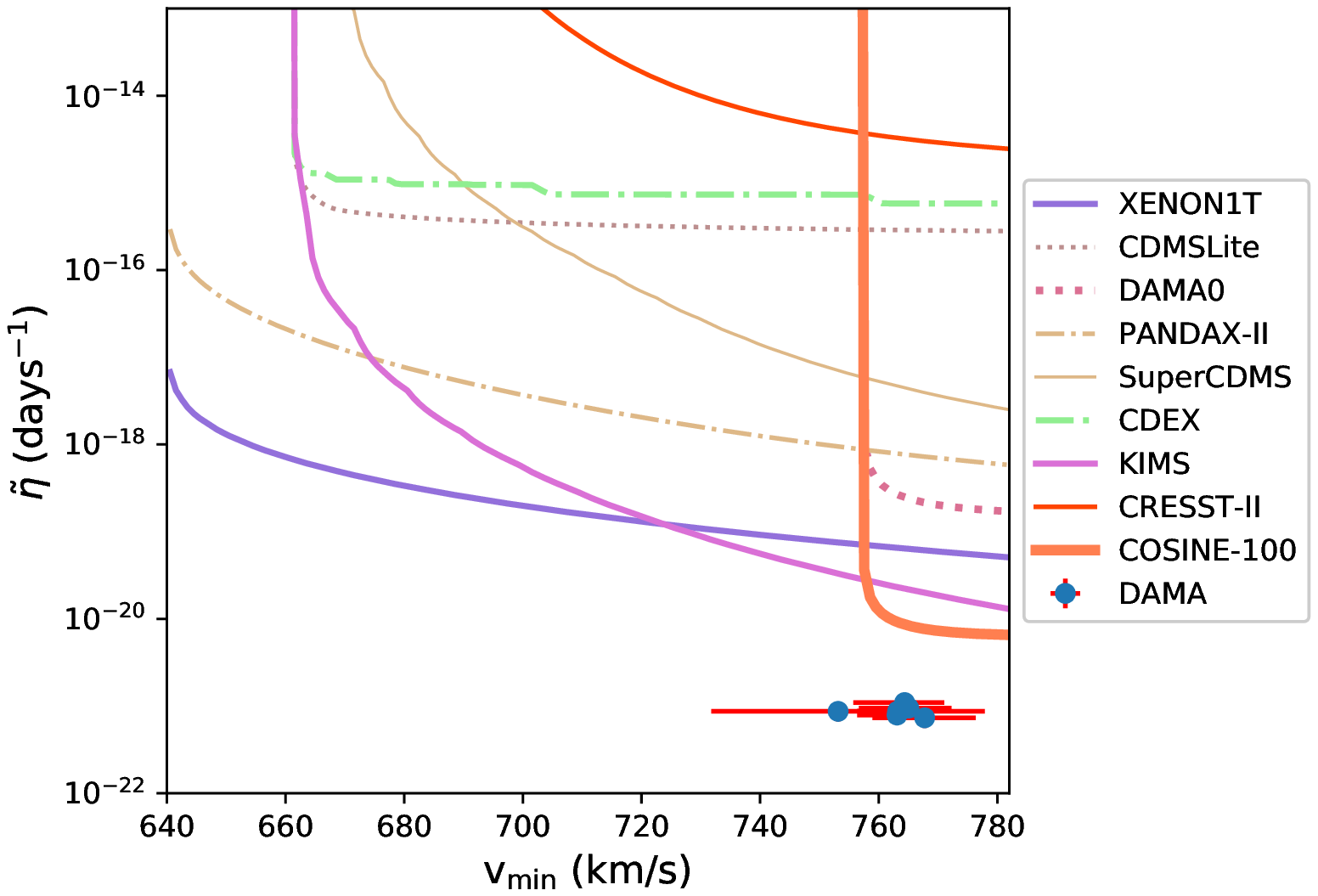}
\end{center}
\caption{(a)~The 5--$\sigma$ best-fit DAMA region for the pSIDM scenario
  is compared to the corresponding 90\% C.L. upper bounds from other
  DM searches for a Maxwellian WIMP velocity distribution and the IDM mass splitting $\delta=18.3$ keV. (b)~Measurements of 
  $\overline{\tilde{\eta}^1}_{[v_{{\rm min},1},v_{{\rm min},2}]}$ (DAMA/LIBRA) and upper bounds
  $\tilde{\eta}^{\rm lim}$ for pSIDM in the
  benchmark point $m_{\chi}$=11.4 GeV, $\delta$=23.7 keV.}
\label{fig:psidm_exclusion}
\end{figure}

In the considered pSIDM scenario, we perform a $\chi^2$ analysis constructing the quantity,
\begin{equation}
\chi^2(m_{\chi},\delta,\sigma_0)=\sum_{k=1}^{14} \frac{\left [S_{{m},k}(m_{\chi},\delta,\sigma_0)-S^{\rm exp}_{{m},k} \right ]^2}{\sigma_k^2},
  \label{eq:chi2}
  \end{equation}
and minimize it as a function of $(m_{\chi},\delta,\sigma_0)$. In the equation above, $S_{m,k}$ and $\sigma_k$ represent the modulation
amplitudes and error measured by DAMA whereas $S^{\rm exp}_{{m},k}$ represents the expected modulation rate. Considering 
a standard isotropic Maxwellian velocity distribution for WIMPs, in Fig.~\ref{fig:psidm_exclusion}(a), the
pSIDM scenario is compared to the corresponding 90\% C.L. upper bounds from other DM searches~\cite{Kang:2018zld}. In this plot, the parameters $\delta=18.3$ keV, $m_\chi=12.1$ GeV and $\sigma_0=7.95\times 10^{-35}$ cm$^2$   
correspond to the absolute minimum of $\chi^2~(\chi^2_{min}=13.19)$. Clearly, the DAMA effect is in strong tension with the upper bounds from PICO-60, KIMS and PICASSO. Interestingly, COSINE-100~\cite{Adhikari:2018ljm} that uses the same $NaI$ target as DAMA does not exclude the pSIDM scenario. Basically, the large modulation fraction in pSIDM in comparison to the elastic case is the reason behind that. We estimated the ratio of the modulation amplitude to the time averaged amplitude,
$S_m^{DAMA}/S_0^{DAMA}$ = $S_m^{DAMA}/S_0^{COSINE}\times S_0^{COSINE}/S_0^{DAMA}$ $\gtrsim$ 0.12,
including a factor $S_0^{COSINE}/S_0^{DAMA}\simeq$ 0.8 due to a
difference between the energy resolutions and efficiencies in the two
experiments, $S_m^{DAMA}\simeq 0.02$ events/kg/day/keVee~\cite{Bernabei:2018yyw}, and $S_0^{COSINE}\lesssim 0.13$ events/kg/day/keVee. In the pSIDM scenario, we found $S_m^{DAMA}/S_0^{DAMA}\gtrsim 0.40$ which explains why COSINE-100 does not constrain pSIDM. The large modulation fraction in 
non-relativistic effective models~\cite{Fitzpatrick:2012ix}, is also the main reason that 
COSINE-100 experiment can not 
rule out all the effective operators allowed by Galilean invariance~\cite{Kang:2019fvz}.  

We further extended our analysis to the halo-independent approach in which the expected rate in a direct detection 
experiment is given by, 
\be
\label{eq:R3}
R_{[\Ed_1, \Ed_2]}(t) = \int_0^\infty \ud\vmin \, \tilde{\eta}(\vmin, t) \,  \eR_{[\Ed_1, \Ed_2]}(\vmin) \, ,
\ee
where $\tilde{\eta}(\vmin, t)$ is the halo function containing the dependence on astrophysics
and $\eR_{[\Ed_1, \Ed_2]}(\vmin)$ is the response function.
Due to the revolution of the Earth around the Sun, the velocity
integral $\tilde{\eta}(\vmin, t)$ shows an annual modulation that can
be approximated by the first terms of a harmonic series,
\begin{equation}
\label{etat}
\tilde{\eta}(\vmin, t) = \tilde{\eta}^0(\vmin) +
\tilde{\eta}^1(\vmin) \, \cos\!\left[ \omega (t - t_0) \right],
\end{equation}

\noindent with the necessary requirement of
$|\tilde{\eta}^1|\le\tilde{\eta}^0$. It is possible to obtain the averages
$\overline{\tilde{\eta}^i}_{[v_{{\rm min},1}, v_{{\rm min},2}]}$
($i=0,1$) directly from the experimental
data $R^i_{[\Ed_1, \Ed_2]}$ as~\cite{DelNobile:2013cva},

\begin{eqnarray}
\overline{\tilde{\eta}^1}_{[v_{{\rm min},1},v_{{\rm min},2}]} 
= \frac{R^i_{[\Ed_1, \Ed_2]}}{\int_0^\infty \ud\vmin \, \eR_{[\Ed_1, \Ed_2]}(\vmin)},
\label{eq:eta_average}
\end{eqnarray}
where the velocity intervals $[v_{{\rm min},1}, v_{{\rm min},2}]$ are
defined as those where the response function
$\eR_{[\Ed_1, \Ed_2]}(\vmin)$ is sizeably different from zero.
In Fig.~\ref{fig:psidm_exclusion}(b), the result for such procedure is shown
for the DAMA/LIBRA--phase2 data with error bars for the
benchmark point $m_{\chi}$=11.4 GeV and $\delta$=23.7 keV. Similarly following~\cite{DelNobile:2013cva}, the upper bound 
on $\tilde{\eta}^{\,0}$ from upper limits on $R_{[\Ed_1, \Ed_2]}^{\rm lim}$ is computed as,
\begin{equation}
\tilde{\eta}^{\rm lim}(v_0) = \frac{R^{\rm lim}_{[\Ed_1,
      \Ed_2]}}{\int_0^{v_0} \ud \vmin \, \eR_{[\Ed_1, \Ed_2]}(\vmin)}\ .
\label{eq:etalim}
\end{equation}
The corresponding upper limits at 90\% C.L. are shown as continuous
lines in Fig.~\ref{fig:psidm_exclusion}(b) for the considered benchmark. 
It is clear that
pSIDM cannot be ruled out as an explanation of the DAMA/LIBRA effect
since in all the energy range of the signal one has
$|\overline{\tilde{\eta}^1}_{[v_{{\rm min},1},v_{{\rm
        min},2}]}|\ll\tilde{\eta}^{\rm lim}$.
        
In Ref.~\cite{Kang:2019uuj} we also considered an extension of the present analysis
to the case of the most general Galilean-invariant WIMP-nucleon effective contact 
interaction for a spin 0, 1/2 or 1 WIMP dark matter following the approach introduced in~\cite{Catena:2016hoj}
and checked the compatibility of the DAMA effect in an
inelastic  scattering scenario. In particular in that analysis, we also included all possible 
interferences among operators and found that in comparison to the elastic case 
discussed in~\cite{Catena:2016hoj}, inelastic
scattering partially relieves but does not eliminate the existing tension between the 
DAMA effect and the constraints from other experiments, when a Maxwellian velocity distribution
for WIMPs is considered. Interestingly in~\cite{Kang:2019uuj}, a small region of the pSIDM parameter space 
scenario discussed here naturally arises 
for $m_\chi\simeq 10$ GeV and $\delta\gtrsim 20$ keV.
\section{Conclusion}
We have analyzed the scenario of proton-philic spin-dependent inelastic dark matter
(pSIDM) for the observed modulation amplitude by DAMA both considering the standard
Maxwellian velocity distribution for WIMPs and adopting a halo-independent
approach. Due to the lower threshold DAMA/LIBRA-phase2 is now
sensitive to WIMP-iodine interactions at low WIMP masses and so pSIDM can no longer explain
the DAMA effect for a Maxwellian velocity distribution of WIMP remaining
consistent with other direct detection experiments. On the other hand when 
the WIMP velocity distribution
departs from a standard Maxwellian, it is possible to explain the observed modulation
amplitude by DAMA in consistency with the results from other direct detection
experiments. The recent COSINE--100 bound
is naturally evaded in the pSDIM scenario due to its large expected
modulation fractions, because inelastic scattering is sensitive to the
high--speed tail of the velocity distribution.

\end{document}